\documentclass{caosp309}

\usepackage{graphicx}

\usepackage{natbib}
\articleNo{B25}
\pubyear{2024}
\volume{54}
\volnumber{1}
\firstpage{1}
\received{November 1, 2023}
\accepted{???}

\begin{document}

\hauthor{E.W.\,Guenther, L.\,Fossati and P. Kab\'{a}th}

\title{The evaporation of planetary atmospheres}

\author{
        E.W.\,Guenther\inst{1}\orcid{0000-0002-9130-6747}
        \and
        L.\,Fossati\inst{2}\orcid{0000-0003-4426-9530}
         \and
         P. Kab\'{a}th\inst{3}\orcid{0000-0002-1623-5352}
               }

\institute{
           Th\"{u}ringer Landessternwarte Tautenburg\\
           07778 Tautenburg, Sternwarte 5 \\
           Germany
           \email{guenther@tls-tautenburg.de}
           \and
           Space Research Institute, Austrian Academy 
           of Sciences, Schmiedl-strasse 6, 8042 Graz, Austria \\
           \and
           \ondrejov                     
           }

\date{November 1, 2023}

\maketitle

\begin{abstract}
  In recent years the focus of exoplanet research has shifted from the
 mere detection to detailed characterization. Precise measurements of
 the masses and radii of transiting planets have shown that some
 low-mass planets have extended atmospheres while others are bare
 rocks. Hybrid atmospheres consisting of a mixture of Hydrogen and
 large amount of heavy elements have also been detected. A key factor
 in explaining this diversity of planetary atmospheres is the erosion
 by the X-ray and EUV-radiation (XUV) from the host-star. The
 evaporation through XUV-radiation has already been measured for a few
 exoplanets.The apparent weakness of the Ca\,II\,HK and the
 Mg\,II\,h\&k emission cores has been interpreted as evidence 
 for the  evaporation of planetary atmospheres.
 The interpretation is that the
 evaporating material from the planet forms a thick torus which
 absorbs the Ca\,II\,HK and the Mg\,II\,h\&k lines from the host
 star. In this contribution  a new way how to prove, or
 disprove this hypothesis by observations is proposed. It is
 furthermore shown that there are enough bright targets already known 
 that can be observed, and more will be found with the PLATO mission.  
\end{abstract}

%

\label{intr}

\section{The evolution of the atmospheres of low-mass planets}
\label{sec1}

Before exoplanets were
discovered, it was generally thought that they would resemble the
planets in our solar system, but research in the past years  have
shown that exoplanets are much more diverse. Precise measurements of
the masses and radii of transiting planets allowed us to gain more
insight into what these planets are. Virtually all Neptunes, Jupiters
and super-Jupiters turned out to have extended hydrogen-dominated
atmospheres. Indeed, it appears that seemingly all planets larger than
1.8 $\rm R_{Earth}$ host extended, hydrogen-dominated atmospheres, but there
are significant differences between Jupiter-mass planets and planets
close to the border of the super-Earth class. 

While Jupiter-mass planets are gas-giants for which much of the mass
is contained inside the Hydrogen/Helium dominated envelope, for the
so-called mini-Neptunes, instead, the Hydrogen/Helium dominated
envelope contains just 1-2\% of the total mass of the
planet. Mini-Neptunes are thus not gaseous planets, but rocky planets
with extended envelopes. In a few cases it was shown that
the atmospheres of mini-Neptunes are hybrid atmosphere containing hydrogen 
and a substantial amount of elements heavier than Hydrogen and Helium 
(Garc\'{\i}a Mu\~{n}oz et al. 2021). The density measurements of low-mass 
planets imply that there are three populations: rocky, water-rich, and gas-rich
planets (Luque \& Pall\'e 2022). 

Close-in planets (a$<$0.1 AU) with radii smaller than 1.4 $\rm
R_{Earth}$ are rocky without an extended atmosphere and close-in
planets with radii larger than 1.8 $R_{Earth}$ are
mini-Neptunes. Close-in planets with radii between 1.4 and 1.8 $\rm
R_{Earth}$ are rare (Fulton \& Petigurea 2018; Owen \& Wu, 2013; Jin
et al., 2014; Lopez \& Fortney, 2014; Fridlund et
al. 2020). Surprisingly, the masses of super-Earths are almost the
same as for mini-Neptunes. These results are very surprising. Why do
some low-mass planets have extended hydrogen atmospheres, and others
not? Furthermore mini-Neptunes and super-Earths are the most common
type of planet, but we do not have any such planet in the
solar-system.

The fact that some of these planets have extended atmospheres, while
others do not, must be related to their formation and evolution
history. Several possible mechanisms have been proposed.
One process is atmospheric erosion (e.g. Lammer et al. in 2014). 
Another possibility is gas-poor formation (e.g. Owen \& Wu, 2013; Lee et al. 2022). 
More recently, Venturini et al. (2020) presented a model in which 
a planet grows from a moon-mass embryo by either silicate or icy pebble accretion,
followed by type I-II migration, and photoevaporation driven mass-loss.
Atmospheric erosion could either be
caused by photoevaporation due to the X-ray and EUV (together XUV)
irradiation from the host-star, or by atmospheric escape due to
formation heating (i.e. core powered mass-loss; e.g. Izidoro et
al. 2022). However, even if gas-poor formation or formation heating
were the dominant processes, atmospheric erosion due to the XUV
irradiation from the host star would still play an important role.
Tian \& Heng (2023) have shown that hybrid atmospheres 
are a natural outcome of the evolution in atmospheres of close-in, low-mass 
planets. 

\begin{figure}
\centerline{\includegraphics[width=0.90\textwidth,angle=0.0]{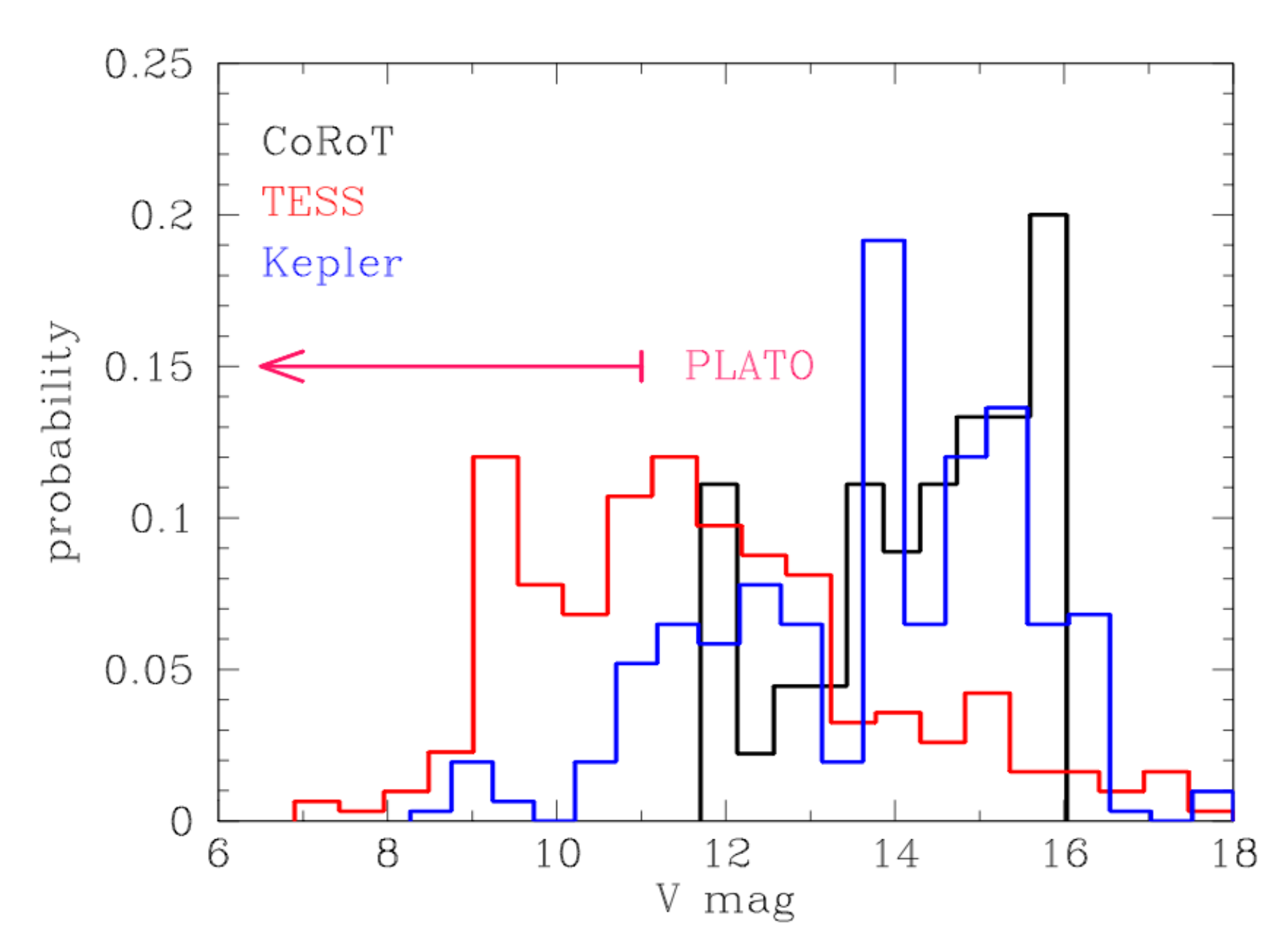}}
\caption{Brightness in V of the planet host stars discovered by CoRoT, Kepler and TESS. The P1 sample of PLATO
contains host stars between V=4 and V=11.}
\label{all}
\end{figure}

\section{Evidence for mass-loss}
\label{sec2}

Arguably, the best studied cases for atmospheric erosion are the hot Jupiters
HD 209\-458 b and HD 189733 b. The detection of atomic hydrogen beyond the
Roche lobe first established that atmospheric material was escaping
from HD 209458 b. Estimated escape rates for HD\,209458\,b are of the order
of $10^{10}$-$10^{11}$ $\rm g\,s^{-1}$ (e.g. Vidal-Madjar et al. 2003,
Garc\'{\i}a Mu\~{n}oz 2007) and $\rm 10^{9}$ - $10^{11}$ $\rm g\,s^{-1}$ for
HD189733b (e.g. Lecavelier Des Etangs et al. 2010).  Therefore, both
hot Jupiters loose about 0.1 to 0.2\% of their mass per Gyr. Thus, the
observed atmospheric escape of the gas-giants is not large enough to
fully erode these planets to their rocky cores.

Since the Hydrogen/Helium atmosphere of a mini-Neptune contains only
1-2\% of the total mass of the planet, mass loss becomes critically
important for low-mass planets.  Since the
amount of XUV radiation of a solar-like star with an age of 10 Myrs is
300 to 1500 times higher than that of the current Sun, the
main erosion phase occurs when the star is young (e.g. Ketzer \&
Poppenhaeger 2023). Therefore, it is possible that the Hydrogen/Helium
atmosphere of a min-Neptune is completely eroded in the first few
Myrs. For example, V1298 Tau is a young star with four transiting
exoplanets for which Poppenhaeger et al. (2021) showed that the
innermost two planets may lose their hydrogen-dominated atmospheres to
become rocky planets. Another interesting case is K2-33b, a young
planet in Upper Sco, which may lose its entire hydrogen atmosphere if
the planet has less than 7-10 $\rm M_{Earth}$ (Kubyshkina et
al. 2018). HST observations of the Neptune-sized planet GJ3470b show
an evaporation rate $\rm 10^{10} g\,s^{-1}$. 
The planet may have already lost
about 35\% of its mass (Bourrier et al. 2018). 
Kepler-1520\,b (=KIC 12557548\,b) 
may have lost $\sim$ 70\% of its formation mass, today we may 
be observing its naked iron core (Perez-Becker \& Chiang  2013).
However, the erosion rates of 
planets with hybrid atmospheres are lower than those of pure Hydrogen-dominated 
atmospheres (Tian \& Heng 2023). 

\section{The observing strategy}
\label{sec3}

Mass loss certainly is a key factor in the evolution of a
planet.  In the case of WASP-12\,b Haswell et al. (2012) detected the 
gas escaping from this heavily irradiated planet. They
observed the planet in the UV and found that the transit is three
times deeper than in the optical, indicating the presence of diffuse
gas, extending well beyond the Roche lobe. They also found that
surprisingly the MgII h\&k line cores have zero flux indicating that
the inner portions of these strong resonance lines are likely affected
by extrinsic absorption due to the material escaping from the planet
(Fossati et al. 2013). There is thus strong evidence for mass loss
for WASP-12b. 

Mass loss from a close-in planet can form a diffuse circumstellar gas
cloud, which absorbs in the cores of strong resonance lines (e.g. CaII
H\&K and MgII h\&k) seemingly suppressing the signatures of stellar
activity below their true value (Haswell et al. 2012; Fossati et
al. 2013).  Simulations show that the density of the material in the
torus is inhomogeneous (Bell et al. 2019; Dwivedi et al. 2019; Zhang
et al. 2023).  Observations of the Helium triplet line at 1083.3 nm
during roughly a quarter of the orbit by Zhang et al. et al. (2023)
indicate the material forms a torus.

Since the density of the material in tours varies along the orbit of the 
planet, the strength of the CaII H\&K lines should also vary.
By observing several orbits, it is then possible to reconstruct the 
density of material in the torus.  However, if we observe only the 
CaII H\&K lines, we would only constrain how much Ca has escaped from the
planet. The  strength of the absorption would depend on the amount of
Ca that the envelope contains.
Using not only the CaII H\&K lines but 
also H$\alpha$ and NaD it would be possible to determine the physical 
properties of the torus, for example, how much mass it
contains. In this way it would be possible to "see" the material that
is escaping from the planet.

Shkolnik et al. (2003, 2005, 2008) have studied the chromospheric 
activity in the Ca II H and K lines with the aim to search for induced
stellar activity. A spectralline that originates from the star has width
that is of the order of the v\,sin\,i of the host star, depending on the 
location of the active region on the stellar surface. The RV-amplitude
of a close-in planet is much larger than that. A ring of material the 
originates from material that has escaped from a close-in planet will 
have a width that corresponds to the RV-amplitude of the planet.
Such a feature will thus be much broader than a stellar spectralline. 

\section{Preparing for PLATO and planed observations}
\label{sec4}

The question is  whether there are already good targets for such a project.
Because the absolute mass-loss of gas-giants are larger than that of
min-Neptunes, gas-giants are the preferred targets.
Table\, \ref{tab:planets} gives a list of nineteen known planets in the mass-range
between $\rm Mp=0.5$ and 5.0 $\rm M_{Jup}$ with $\rm Rp\geq1.4\,R_{Jup}$
orbiting stars with $\rm V\leq 11\,mag$
that are observable from La Silla
\footnote{Objects selected using  \url{https://https://exoplanet.eu}}.
As can be seen, all of these targets are short-period planets. The large radii are thus 
most likely due to large heating by the host star. Thus, these planets are likely to have also a large
mass-loss rate. For comparison WASP-12\,b has $\rm Rp=$ 1.47 $\rm M_{Jup}$, 
$\rm Mp=1.9$ $\rm M_{Jup}$ (Collins et al. 2017).

While many bright stars hosting close-in gas-giant have
already been identified it
would be even more exciting if bright stars hosting
close-in Neptunes were found. Absolutely thrilling would be the discovery of
planets like those of V1298 Tau
(V=10.1 mag), GJ3470b (G=11.4 mag), or K2-33b (G=14.1 mag), 
K2-240  (V=13.4 mag) but orbiting brighter stars.

What are the prospects to find planets like that but orbiting brighter stars?  
The key to find these is the PLATO mission. 
PLATO (PLAnetary Transits and
Oscillation of stars) will search for extrasolar planets by means
of ultra-high-precision transit photometry. PLATO consists of 24
normal and 2 two high cadence cameras providing combined wide field
of view (FoV) of 49x49 square degrees.  The PLATO mission and the 
key-science goals are summarized in Rauer et al. (2014, 2021).

The first PLATO field has now been selected
\footnote{See \url{https://www.cosmos.esa.int/web/plato/first-sky-field}}.
The center is at RA:
06:21:14.5 DE: -47:53:13. 
This field contains more than 9000 dwarf and
subgiant stars of spectral types from F5 to K7 with V $< $11 mag (P1
sample) that will be observed. The random noise of the P1-sample will be 
lower than 50 ppm in one hour.  Fig.\,\ref{all} shows the
brightness of planet host stars discovered in the CoRoT, Kepler and
TESS mission, respectively \footnote{Figures produced using  \url{https://https://exoplanet.eu}}.
AS can be seen in the figure, one of the advantages of PLATO compared with previous
missions is that planet host stars will be much brighter.  The 
baseline observing strategy assumes that this
field will be observed for at least two years. The PLATO spacecraft
has been designed to perform scientific operations for at least 8.5
years. PLATO will also observe more than 159,000 dwarf and subgiant stars of
spectral types from F5 to K7 with mV $<$ 13.

Ondrejov observatory, the Th\"uringer Landessternwarte Tautenburg and
the Pontificia Universidad Cat\'olica de Chile as main partners have a 
unique access  to the ESO1.5m telescope in La Silla (Chile). 
Minor partners are the IGAM of the university of Graz,  Universidad Adolfo
Ibanez and Masaryk University. 
The telescope has recently been refurbished.  
It is fully operational and currently equipped with the PUCHEROS Echelle
spectrograph that has a resolution of R=20,000.  By the end of 2024,
we will install PLATOspec, which is a state-of-the-art UV-optimised
high-resolution Echelle spectrograph with a resolution of R=68,000
covering the 360-680 nm spectral range without gaps. The high UV
throughput of the spectrograph, which is rather unique in the current
instrumental landscape, enables one to collect high-quality spectra
throughout the entire optical band, including the CaII H\&K lines, the
NaI D lines, and the H$\alpha$ line. The spectrograph will be fibre fed and
will be placed in a temperature-stabilized room.

As the name of the instrument already suggest, its main purpose will
be the follow-up observations of transiting planets discovered in the
PLATO mission. Because of the relatively large amount of observing time
required this instrument is ideal to carry the project.

\begin{table*}
\tabcolsep=2.5pt
\small
\caption{Inflated gas-giant planets observable from La Silla orbiting stars brighter than 11 mag in V.}
\begin{tabular}{l l c c l l l l}
\hline\hline
Star &  RA{\tiny 2000}    &  DE{\tiny 2000}     &  V  &  Mass  &  Radius  &  Period  &  K \\
     & {\tiny (hh:mm:ss)} &  {\tiny (dd:mm:ss)} & [mag]  &  $\rm [M_{Jup}]$  & $\rm [R_{Jup}]$  & [days]  & [m/s] \\
\hline
HIP65A        & 00:00:45  &  -54:49:51  &  11.1    &  $3.213\pm0.078$           &  $2.03_{-0.49}^{+0.61}$     &  0.98 &  754 \\
HD2685       & 00:29:19  &  -76:18:15  &  9.6     &  $1.18\pm0.09$              &  $1.44\pm0.01$                    &  4.13   &  118	\\
WASP-76     & 01:46:32  &  +02:42:02  &  9.5   &  $0.92\pm0.03$                &  $1.83\pm0.06$                  &  1.81  &  119 \\
WASP-79     & 04:25:29  &  -30:36:02  &  10.1  &  $0.9\pm-0.09$     &  $1.7\pm0.11$     &  3.66  &   88 \\
WASP-100   & 04:35:50  &  -64:01:37  &  10.8  &  $2.03\pm0.12$    &  $1.69\pm-0.29$    &  2.85   &   215 \\
WASP-82     & 04:50:39  &  +01:53:38  &  10.1  &  $1.24\pm0.04$      &  $1.67_{-0.05}^{+0.07}$   &  2.71   &  130 \\
TOI-640       & 06:38:56  &  -36:38:46  &  10.5   &  $0.88\pm0.16$             &  $1.771\pm0.06$            &  5.00  &  78 \\
WASP-121   & 07:10:24  &  -39:05:51  &  10.4  &  $1.184_{-0.064}^{+0.065}$ &  $1.865\pm0.044$ &  1.27  &  181 \\
KELT-17       & 08:22:28  &  +13:44:07  &  9.6   &  $1.31\pm0.29$    &  $1.525_{-0.06}^{+0.065}$   &  3.08  &  131 \\
HAT-P-69     & 08:42:01  &  +03:42:38  &  9.8   &  $3.58\pm0.58$             &  $1.676_{-0.033}^{+0.051}$ &  4.79  &  309 \\
TOI-2669      & 08:58:53  &  -13:18:45  &  9.5   &  $0.61\pm0.19$             &  $1.76\pm0.16$             &  6.20     &  60 \\
HD85628A    & 09:50:19  &  -66:06:50  &  8.6   &  $3.1\pm0.9$               &  $1.53_{-0.04}^{+0.07}$    &  2.82    &  307 \\
WASP-15      & 13:55:43  &  -32:09:35  &  10.9  &  $0.542\pm0.05$   &  $1.428\pm0.077$ &  3.75  & 63  \\
NGTS-2        & 14:20:30  &  -31:12:07  &  11.0  &  $0.67\pm0.089$            &  $1.536\pm0.062$           &  4.51  &  69 \\
WASP-189    & 15:02:44  &  -03:01:53  &  6.6   &  $1.99_{-0.14}^{+0.16}$    &  $1.619\pm0.021$           &  2.72   &  182 \\
WASP-88      & 20:38:03  &  -47:32:17  &  10.4 &  $0.56\pm0.08$    &  $1.7_{-0.07}^{+0.13}$      &  4.95   &  57  \\
MASCARA-1 & 21:10:12  &  +10:44:20  &  8.3  &  $3.7\pm0.9$       &  $1.5\pm0.3$       &  2.15    &  405 \\
HD202772A   & 21:18:48  &  -26:36:59  &  8.3  &  $1.008_{-0.079}^{+0.074}$ &  $1.562_{-0.069}^{+0.053}$ &  3.31    &  97 \\
WASP-11      & 21:55:04  &  -22:36:45  &  10.3  &  $1.85\pm0.16$    &  $1.442\pm0.094$ &  2.31    &  212 \\
\hline\hline
\end{tabular}
\label{tab:planets}
\end{table*}

\acknowledgements
This work was generously supported by the Th\"uringer Ministerium f\"ur Wirtschaft, Wissenschaft und Digitale Gesellschaft.  
This work has made use of the https://exoplanet.eu. We are very grateful  to the  
{\em exoplanet team} that is providing this service. 
This research has made use of the SIMBAD database, operated at CDS, Strasbourg, France.

\end{document}